\documentclass[aps,prl,twocolumn,superscriptaddress,notitlepage,10pt]{revtex4-1}
\usepackage{amsmath,amssymb,amsfonts}
\usepackage{graphicx}
\usepackage[dvipsnames]{xcolor}
\usepackage{comment}
\usepackage[overload]{empheq}
\usepackage{upgreek}
\usepackage{lmodern,pdftexcmds}
\usepackage{hyperref}
\usepackage{soul}
\usepackage[normalem]{ulem}

\hypersetup{
    bookmarksopen=true,
    unicode=false,          % non-Latin characters in Acrobat’s bookmarks
    pdftoolbar=true,        % show Acrobat’s toolbar?
    pdfmenubar=true,        % show Acrobat’s menu?
    pdffitwindow=false,     % window fit to page when opened
    pdfstartview={Fit},    % fits the width of the page to the window
    pdftitle={},    % title
    pdfauthor={},     % author
    colorlinks=true,       % false: boxed links; true: colored links
    linkcolor=black,          % color of internal links
    citecolor=black,        % color of links to bibliography
    urlcolor=black,           % color of external links
    pdfborder={0 0 0}
}

\newcommand{\beq}{\begin{equation}}
\newcommand{\eeq}{\end{equation}}
\newcommand{\beqa}{\begin{eqnarray}}
\newcommand{\eeqa}{\end{eqnarray}}

\newcommand\F{\mathcal{F}}

\renewcommand{\bm}[1]{\boldsymbol{\mathbf{#1}}}

\providecommand*{\dd}[3][]{\frac{\mathrm{d}^{#1}#2}{\mathrm{d} #3^{#1}}}

\newcommand{\qmbox}[1]{\quad \mbox{#1} \quad}

\providecommand*{\rme}{\mathrm{e}}
\providecommand*{\rmd}{\mathrm{d}}
\providecommand*{\rmi}{\mathrm{i}}

\providecommand*{\bm}{\mathbf}
\providecommand*{\m}{\displaystyle}

\renewcommand{\F}{\mathcal{F}}

\begin{document}

\preprint{}

% \title{Surfing its own wave: hydroelastic migration of a particle near a membrane}
\title{Curvature regularization near contacts with stretched elastic tubes}



\author{Bhargav Rallabandi}
\email{bhargav@engr.ucr.edu}
\affiliation{Department of Mechanical Engineering, University of California, Riverside, California 92521, USA }
\affiliation{Department of Mechanical and Aerospace Engineering, Princeton University, Princeton, New Jersey 08544, USA}
%\thanks{B.R and J.M. contributed equally to this work.}

\author{Joel Marthelot}
%\email{}
\affiliation{Department of Chemical and Biological Engineering, Princeton University, Princeton, New Jersey 08544, USA}
%\thanks{B.R and J.M. contributed equally to this work.}

\author{P.-T. Brun}
%\email{}
\affiliation{Department of Chemical and Biological Engineering, Princeton University, Princeton, New Jersey 08544, USA}

\author{Jens Eggers}
\email{majge@bristol.ac.uk \\}
\affiliation{School of Mathematics, University of Bristol, University Walk, Bristol BS8 1 TW, UK}

\date{\today}

\begin{abstract}
Bringing a rigid object into contact with a soft elastic tube causes the tube to conform to the surface of the object, resulting in contact lines. The curvature of the tube walls near these contact lines is often large and is typically regularized by the finite bending rigidity of the tube. Here, we show using experiments and a F\"{o}ppl--von K\'{a}rm\'{a}n-like theory that a second mechanism of curvature regularization occurs when the tube is axially stretched. The radius of curvature obtained is unrelated to the bending rigidity of the tube walls, increases with the applied stretching force and decreases with sheet thickness, in contrast with the effects of finite bending rigidity. %Moreover, the axial force decreases the contact area between the tube and the intruding object, potentially reducing the drag necessary to propel the object through the tube. 
We show that these features are due to an interplay between geometry and mechanics specific to elastic tubes, but one that is absent from both planar sheets and spherical shells. 
\end{abstract}

%\keywords{}

\maketitle 

The mechanics of thin sheets (or plates) is determined by the
interplay between
bending and stretching \cite{Witten07,CM03,TV17}. As the sheet is deformed
from its rest state, it bends out of its plane, but it must also
stretch in its own plane, in order to accommodate its new shape. Bending
rigidity $B$ prevents the sheet from turning too sharp a corner, so it is
the ratio of $B$ to some other scale that sets the size of the smallest
ridge \cite{Witten07}, or the wavelength of wrinkles \cite{CM03,TV17}. 
Shells are sheets that possess a non-trivial rest shape, such as a
sphere or a cylinder. This introduces a non-trivial differential
geometry into the problem, and introduces an extra length scale, such as
the radius of the undeformed cylinder.

\begin{figure} 
    \includegraphics[scale=0.5725]{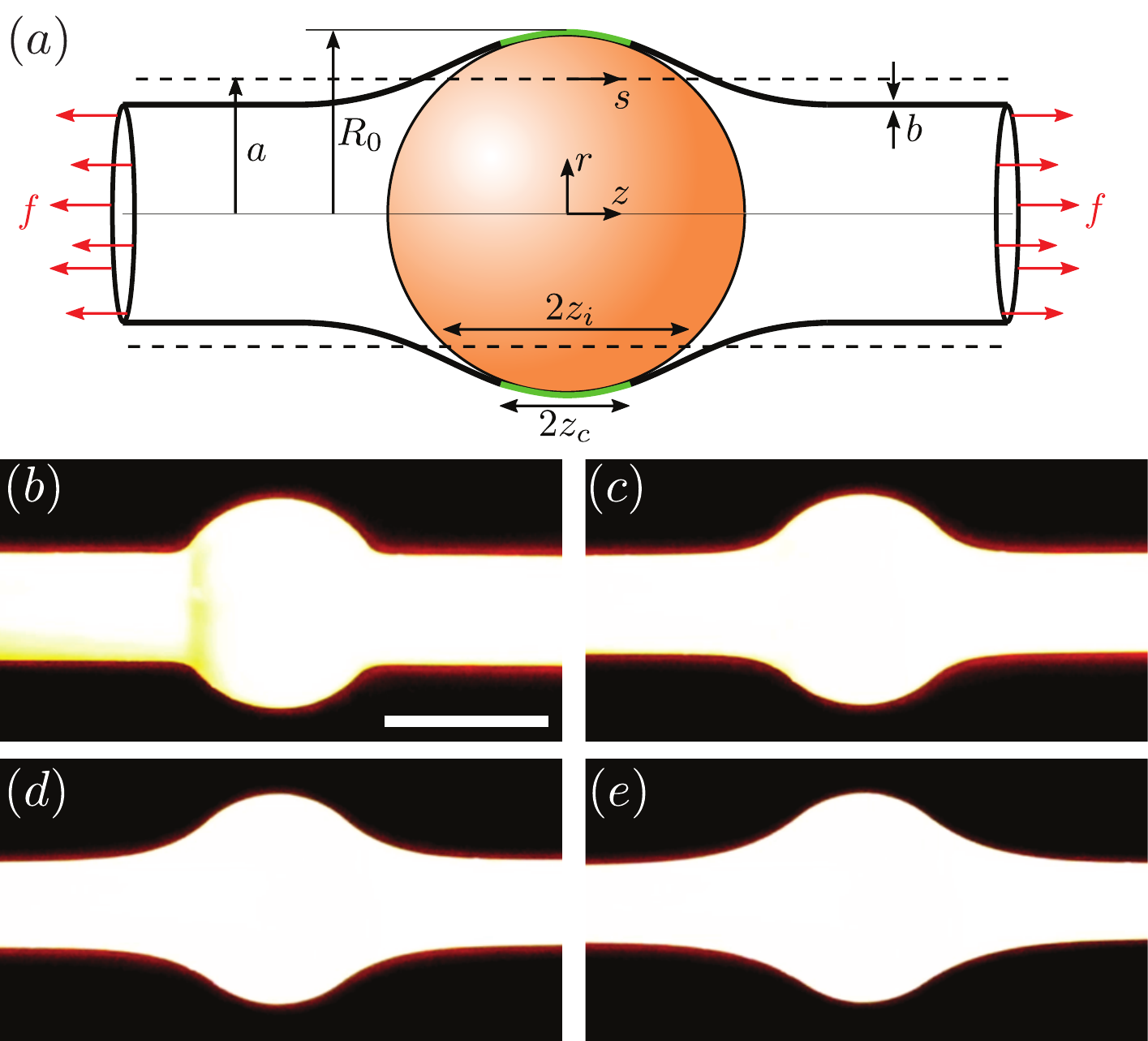}
    \caption{(a) Sketch of the setup showing an axisymmetric intruder in a tube (solid lines) that is cylindrical (radius $a$) in its undeformed reference configuration (dashed lines). The tube is stretched by applying equal and opposite forces per length of magnitude $f$ at the tube ends.  The arclength coordinate
$s$ in the undeformed configuration is a material label. The intersection of the undeformed tube and the intruder defines a geometric intersection length $2z_{i}$ while the region of contact (highlighted in green) has axial extent $2 z_{c}$.
(b)--(e) Experimental pictures of the shape of the tube of radius $a=3.25$ mm around an intruder of radius $R_0=6.35$ mm for increasing value of $f$. Scale bar is 10 mm.}
\label{FigSetup}
\end{figure}

A particular focus of previous
research has been the behavior of spherical shells; here we focus on
cylindrical shells or tubes, which introduces a high degree of anisotropy
into the problem, as they are curved in only the azimuthal direction, 
but flat along the axis. In particular, an object can be confined inside
a tube without applying any external force, in which case a sharp 
corner forms at the geometrical intersection between the cylinder and
the intruder. Here we will look at various ways in which this corner
is regularized, even without invoking any bending rigidity. The resulting
length scale bears some resemblance with the ``elastic capillary length''
introduced previously in the context of wrinkling \cite{VAVB11,TV17}.

We consider the contact mechanics of a thin-walled elastic tube that
in its undeformed reference configuration has a cylindrical shape with
radius $a$, into which is introduced a larger axisymmetric intruder of maximum radius $R_0 > a$ (Fig.\,\ref{FigSetup}a) \cite{L68,TB11,LM18}.
For simplicity we also assume that the intruder has a symmetry plane, here identified with $s = z = 0$, where $s$ is the axial arclength coordinate in the reference state (Fig.\,\ref{FigSetup}a) and is therefore a material label. The tube may also be subject to a traction (force per area) $\bm{\sigma}(s) = \{ \sigma_r(s), \sigma_z(s)\}$ acting along its surface and forces per length $\bm{f}^i = \{f_r^i(s), f_z^i(s)\}$ acting at the circular rims at the ends of the tube (Fig.\,\ref{FigSetup}a); $\bm{\sigma}$ can either be applied externally or may be a consequence of contact with the intruding object. 

The combined action of the external force and the presence of the intruder deforms the tube walls, resulting generically in contact between the tube and the intruder over a finite region. Figure \,\ref{FigSetup}b--e presents images of a spherical intruder
(304 stainless steel ball bearing, radius $R_0= 6.35$\,mm) in a
commercial cylindrically shaped latex party balloon (rest radius $a=3.25$\,mm,
thickness $b=320\,\mu$m), clamped on one side and attached to the load
cell of a universal testing machine (Instron) on the other side. We
record the tensile force $F$ as the cylindrical tube is stretched.
The shape of the deformed tube is imaged with a high-resolution camera that is mounted on a motorized linear translation stage to follow the intruder during the stretching. 

As can be clearly seen in Figs.\,\ref{FigSetup}b-e, stretching the tube increases the length scale over which the tube relaxes to its cylindrical shape. Simultaneously, the contact area between the soft tube and the rigid intruder decreases. We are interested in the structure and size of the contact region generated by the geometry of the intruder and the applied forces. As we will show, curvature singularities that occur nominally at the contact of the thin elastic tube with the surface of the intruder are not only regularized by the finite bending rigidity of the tube walls, but can be independently controlled through the axial stretching of the tube. %Furthermore, axial stretching also determines the area of contact, and is therefore a useful way to control the mobility of the intruding object inside the tube. 

We model an intruder inside a cylindrical tube
using a nonlinear shell theory
analogous to the F\"{o}ppl--von K\'{a}rm\'{a}n theory for plates
\cite{landau_lifschitz86_elasticity}, applicable for small strains
and moderate rotations of the elastic tube surface. In this theory,
out-of-plane deformations and in-plane stretching are treated within
linear elasticity, but stretching is coupled geometrically to potentially
large deformations of the shell from its rest state, resulting in a strongly
{\it nonlinear} theory. Defining the displacements
$\bm{u}(s) = (u_r(s), u_z(s))$ and following the shell theory developed
in \citet{audoly_book} (p. 447), the deformation of the tube is
governed by the radial and axial stress balances
\begin{subequations} \label{Shell}
\begin{align}
   \label{Shellr}
   &\dd{(N_s \dd{u_r}{s})}{s} - \frac{N_{\theta}}{a} + \sigma_r = 0,  \\
   \label{Shellz}
   &\dd{N_s}{s} + \sigma_z = 0,
\end{align}
\end{subequations}
where $N_s(s)$ and $N_{\theta}(s)$ are, respectively, the axial and
azimuthal diagonal elements of the in-plane stress tensor
(with units of force per length). 
In particular, $N_{\theta}$ is the hoop stress characteristic of
cylindrical geometry. Equation \eqref{Shellr} has the form of a membrane
equation \cite{landau_lifschitz86_elasticity}, where the second term
resembles the capillary pressure term of a fluid cylinder \cite{EV08},
with the tension $N_{\theta}$ playing the role of surface tension. 
Expanding for small wall thickness $b \ll a$
(Young's modulus $E$, Poisson's ratio $\nu$), one finds
$N_s = \frac{Eb}{1-\nu^2}(u_z' + \frac{1}{2} \left(u_r'\right)^2 +
\nu \frac{u_r}{a})$ and $N_{\theta} = \frac{Eb}{1-\nu^2}\left(\frac{u_r}{a} +
\nu\left(u_z' + \frac{1}{2}\left(u_r'\right)^2\right)\right)$,
where primes denote derivatives with respect to the argument
(in this case the undeformed arclength coordinate $s$) \citep{audoly_book}. The bending rigidity of the tube
walls has been neglected in \eqref{Shell}, but will be considered below.
In the limit $a\rightarrow\infty$ one recovers the standard
plate equations, and the hoop stress drops out of the description. We focus on the case relevant to our experiments, viz. no tractions act on the non-contacting part of the tube surface and purely axial stretching forces act at the tube ends ($f_r^i = 0, f_z^i = \pm f$); cf. Fig. \ref{FigSetup}a.  Equilibrium at the ends of the tube sets the boundary condition $N_s(s\rightarrow \infty) = f = F/(2 \pi a)$.

\begin{figure}
\centering
\includegraphics[scale=0.68]{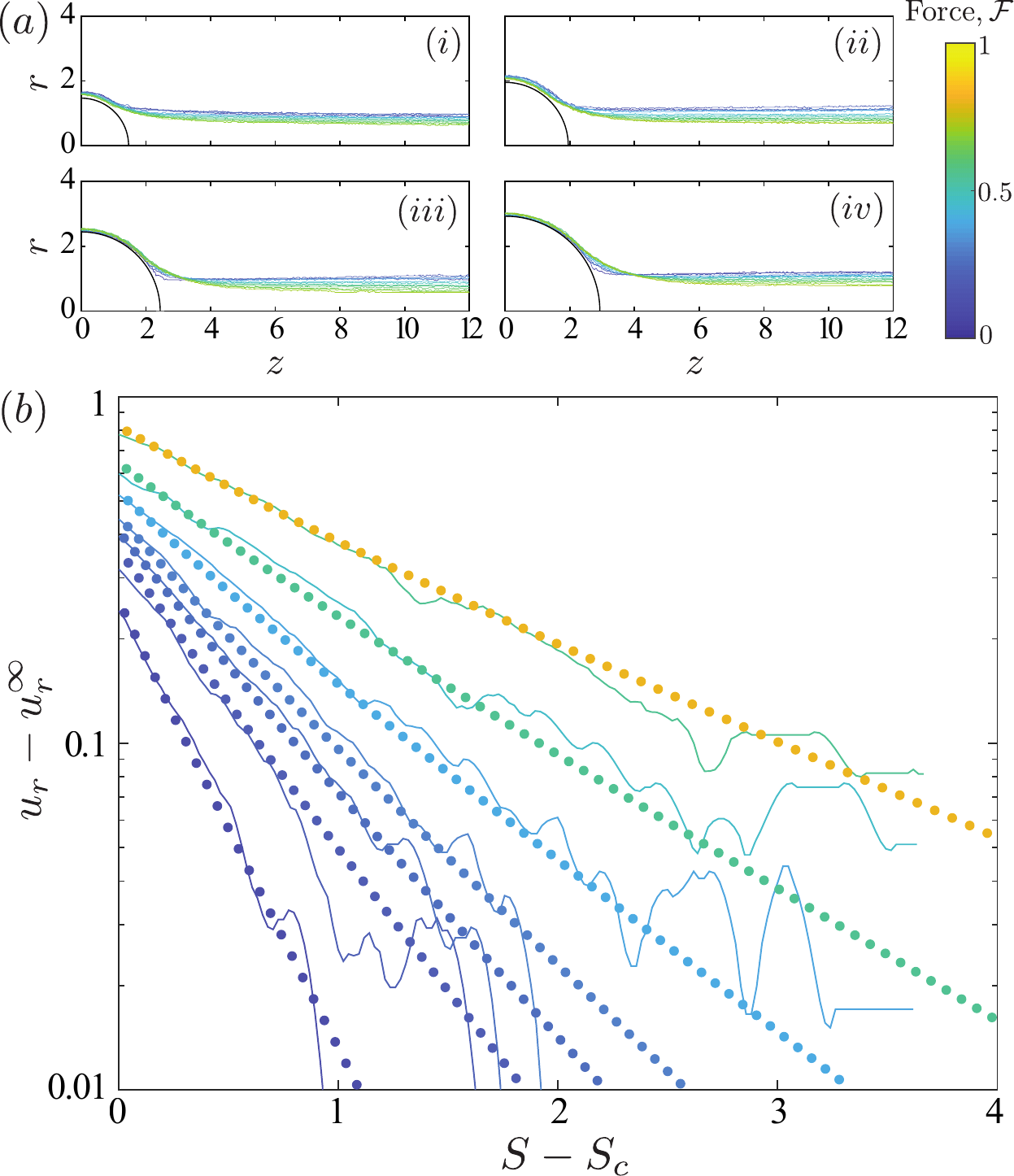}
\caption{(a) Experimental profiles obtained for four different intruders of dimensional radii $R_0/a=$ 1.46 (\textit{i}), 1.95 (\textit{ii}), 2.44 (\textit{iii}), 2.93 (\textit{iv}). The dimensionless force $\mathcal{F}$ is coded in color. (b) Exponential decay of the radial displacement $u_r-u_r^\infty$ with the distance from the contact point $S-S_c$ for increasing dimensionless stretching forces. Solid lines are experiments for $\mathcal{F}=0.13$, 0.22, 0.27, 0.32, 0.38, 0.46 and 0.55, and symbols are theoretical predictions for $\mathcal{F}=0.1$, 0.2, 0.25, 0.3, 0.4, 0.55 and 0.8.}
\label{FigStretch_exp}
\end{figure}

%(a) Numerical prediction of the tube shape for different stretch $ \F$ with $\nu=1/2$  and a spherical intruder with $R_0=1.5$. As $ \F$ increases the length of the contact region 2$z_c$ shrinks, while the length scale over which the tube relaxes to a cylindrical shape increases. The curvature at the contact point diverges as $ \F^{-1/2}$. The dashed line indicates the undeformed shape of the tube. 

Figure \,\ref{FigStretch_exp}a presents the shapes of such tubes obtained experimentally for four intruder diameters. The shapes are color-coded using the value of the dimensionless stretching force $\mathcal{F}=f/(Eb)$ where $E=1.29$\,MPa is the Young modulus of the latex tube.  The shape of the tube depends both on the applied stretching force per circumferential length $f$ and the shape of the intruder, given by $r = R(z)$. In experiments, it is convenient to measure radial displacements as functions of the arclength in the \emph{deformed} configuration $S$ rather than in terms of the undeformed coordinate $s$. The tube loses contact with the intruder at two contact points $s = \pm s_c$ ($S = \pm S_c$) that depend on the stretching force and must be determined as part of the solution. Our experimental data suggest an exponential relaxation of the radial displacement $u_r-u_r^\infty$ with distance from the contact point $S-S_c$ (lines in Fig.\,\ref{FigStretch_exp}b) over a length scale that increases with the dimensionless force. We now turn to predict this relaxation length theoretically.

It is convenient to rescale all lengths by $a$ and define $\tilde{s} = s/a$, $\tilde{s}_c = s_c/a$, $\tilde{\bm{u}}(\tilde{s}) = \bm{u}(s)/a$, $\tilde{R}_0 = R_0/a$ etc.; the contact location $s_c$  is unknown and must be determined as part of the solution. In the following, we focus on $s \geq 0$ using symmetry
and drop tilde accents for convenience.  On the
non-contacting part of the tube surface ($s >s_c$), where $\bm{\sigma} = \bm{0}$,
(\ref{Shell}) reduce to 
\begin{subequations} \label{FreeGE}
\begin{align}
&\mathcal{F} u_r'' - u_r - \nu \mathcal{F} = 0, \\
&u_z' + \frac{1}{2} \left(u_r'\right)^2 + \nu u_r = (1-\nu^2)\mathcal{F},
\end{align}
\end{subequations}
where $\mathcal{F} = f/(Eb) = N_s/(E b)$ is the dimensionless
stretching force introduced previously. Solving \eqref{FreeGE} yields general solutions
\begin{subequations} \label{FreeSol}
\begin{align}
u_r(s > s_c)  &= - \nu\F + C\, \rme^{-(s-s_c)/\sqrt{\F}}, \\
u_z(s > s_c) &= D + (s-s_c)\F  + \nu\sqrt{\F}   C \rme^{-(s-s_c)/\sqrt{\F}} \nonumber \\
&\qquad + \frac{C^2\, \rme^{-2 (s-s_c)/\sqrt{\F}} }{4 \sqrt{\F}}, 
\end{align}
\end{subequations}
where $C$, $D$ and $s_c$ are undetermined constants and we have
dropped exponentially growing solutions. Thus, according to (\ref{FreeSol}a), the tube relaxes
from the contact point over a dimensionless length scale $\sqrt{\F}$,
to a cylindrical shape of radius $1 - \nu \F$. This length scale is
analogous to the ``elastic capillary length'' \cite{VAVB11,TV17}, but where
instead of the intrinsic pressure $N_{\theta}/a$ produced by the applied
force, the pressure is imposed from the outside. 

\begin{figure}
\centering
\includegraphics[scale=0.92]{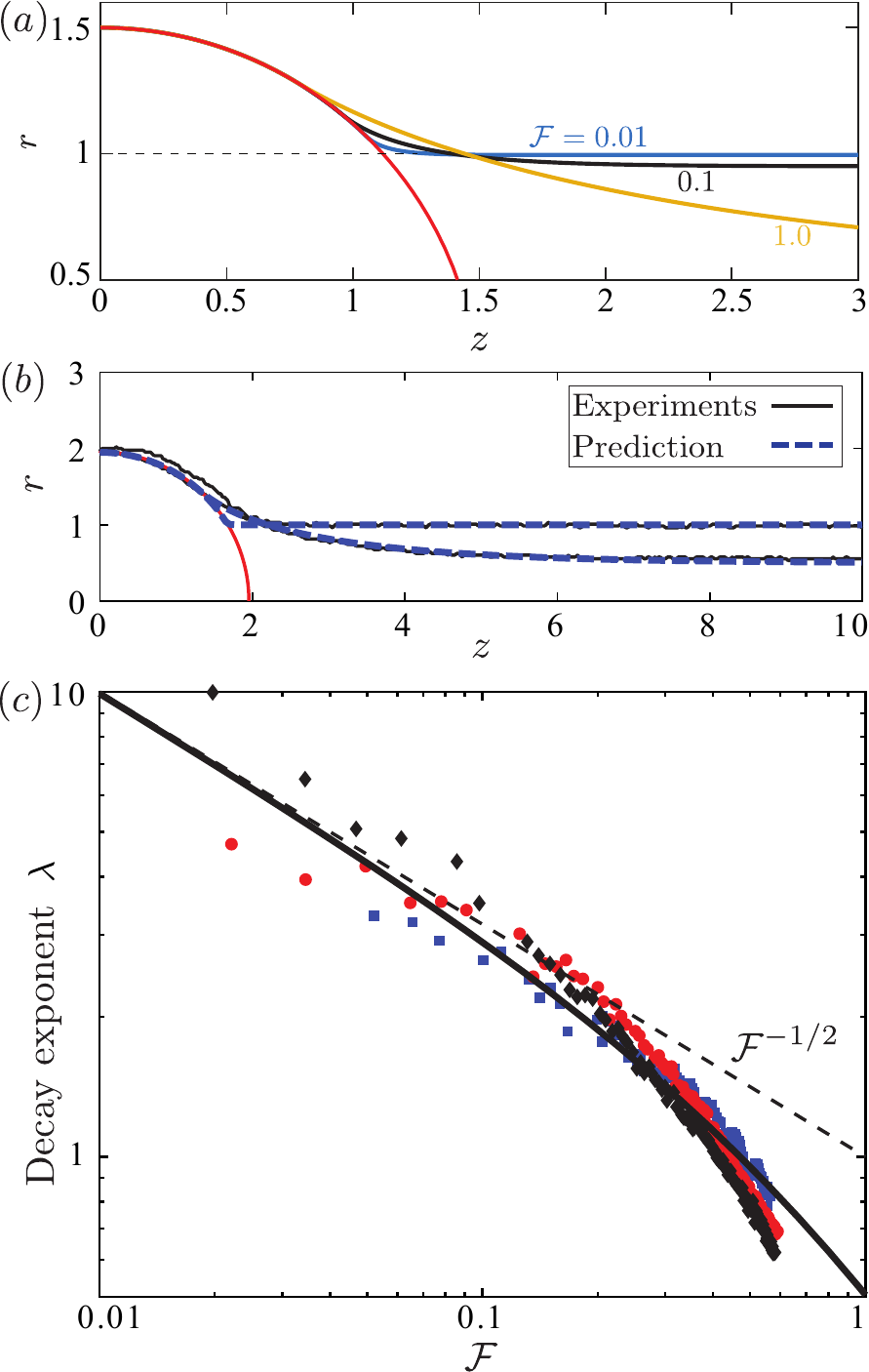}
\caption{(a) Theoretical prediction of the tube shape for different
stretches $\F$ with $\nu = 1/2$ and a spherical intruder with $R_0/a= 1.5$.
As $\F$ increases, the length of the contact region shrinks, while the length scale over which the tube relaxes to a cylindrical shape increases. The curvature at the contact point diverges as $\F^{-1/2}$. The dashed line indicates the undeformed shape of the tube.  (b) Theoretical prediction (in dashed blue lines) and experimental tube shape (in solid black lines) for $R_0/a=$1.95 and a small ($\F$=0.001) and large ($\F$=0.5) dimensionless stretching force. (c) Decay exponent $\lambda$ versus $\F$ for $R_0/a = 1.46$ (blue squares), 1.95 (red circles) and 2.44 (black diamonds). Solid curves are
theoretical predictions, dashed line is the asymptote for small $\F$.}
\label{FigStretch_theory}
\end{figure}

The unknown constants are determined by the solution in the contact region $(0\leq s \leq s_c)$. Since $z(s) = s + u_z(s)$ is the axial coordinate of a material point $s$ in the deformed state, contact is characterized by the geometric condition $R(z(s)) = 1 + u_r(s)$. This yields the radial deformation gradient in the contact region, $u_r'(s) = (1 + u_z'(s))\,R'(s + u_z(s))$. For frictionless contact, the tension $N_s$ remains constant over the contact region [so (\ref{FreeGE}b) remains valid], though friction will modify this in general. Assuming negligible friction we use (\ref{FreeGE}b) to write $u_z'$ in favor of $u_r'$ and $u_r$, letting us recast the geometric constraint above as 
%\begin{subequations}
\begin{align}  \label{ContactGE}
 u_r' &= \frac{-1 + \sqrt{2 R'^2 \left(1 + (1-\nu^2)\F - \nu u_r \right) + 1} }{R'}
 % u_z' &= (1-\nu^2)\F - \frac{(u_r')^2}{2} - \nu u_r, 
\end{align}
%\end{subequations}
valid for $0\leq s\leq s_c$; the argument of $R$ is $s + u_z(s)$.
Thus, we have reduced the problem to
two uncoupled nonlinear ODEs (\ref{FreeGE}b) and \eqref{ContactGE}
for $u_r(s)$ and $u_z(s)$, which we solve numerically.

The ODEs are subject to the symmetry
condition $u_z(0) = 0$ and continuity of the displacement and stress fields around the contact point, which, on using \eqref{FreeSol} are
$u_r(s_c) = -\nu \F + C$, 
$u_r'(s_c) = - \frac{C}{\sqrt{\F}}$, and
$u_z(s_c) = D + C\nu \sqrt{\F} + \frac{C^2}{4 \sqrt{\F}}$. Eliminating $C$ yields the contact condition
\begin{align}\label{ContactCond}
u_r + \sqrt{\F} u_r' + \nu \F = 0 \qmbox{at} s = s_c^-,
\end{align}
which along with (\ref{FreeGE}b) and \eqref{ContactGE} provides enough conditions to evaluate $s_c$, $C$ and $D$. The dimensionless arclength can then be computed from the numerical solution of the displacement field as $S(s) = {\m \int_{0}^s} \sqrt{\left(1 + u_z'(\tau)\right)^2 + \left(u_r'(\tau)\right)^2}\, \rmd \tau$.

\begin{figure}
\centering
\includegraphics[scale=1]{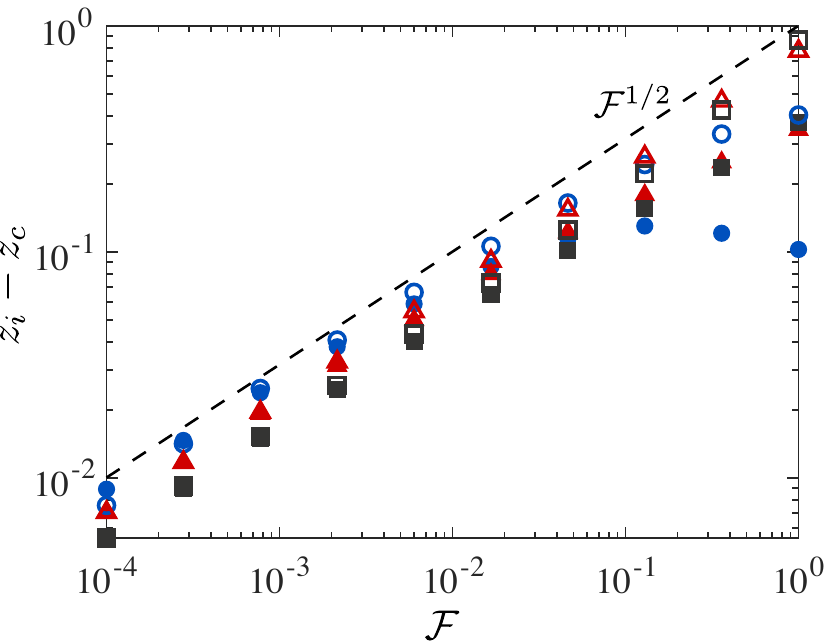}
\caption{Deviation of the contact length $z_c$ from its no-stretch value $z_i$ as a function of the applied stretching force $\F$. Symbols are
numerical results for spherical intruders of different radii [$R_0 = 1.1$ (blue circles); $R_0 = 1.5$ (red triangles); $R_0 = 2.0$ (black squares)] and different Poisson's ratio [$\nu = 0.5$ (filled); $\nu = 0$ (open)]. The dashed line is $\sqrt{\F}$. Point contact at $z_c = 0$ is approached with increasing $\F$.}
\label{FigvsF}
\end{figure}

We solve \eqref{ContactGE} numerically for a spherical $R(z)$ and an applied force $\mathcal{F}$. Typical results are plotted in
Fig.\,\ref{FigStretch_theory}a, which shows both the shape of the tube surface and the location of the contact point for different $\mathcal{F}$. 
As $\F$ decreases, the length $\sqrt{\F}$ over which the tube relaxes [cf. \eqref{FreeSol}] becomes shorter, and the solution relaxes to a cylindrical shape more rapidly from the contact point. In the limit $\F \rightarrow 0$ a corner forms at the ring of intersection between the sphere and the tube $(r = 1,\, z = z_{i} = \sqrt{R_0^2 - 1})$, as shown Fig.\,\ref{FigStretch_theory}a. In general, the contact point moves towards the symmetry point $z = 0$ with increasing $\F$ (Fig. \ref{FigvsF}). In Fig.\,\ref{FigStretch_theory}b we compare the predictions of our theory with experimental profiles for $R_0/a=1.95$ and two dimensionless forces
($\mathcal{F}$=0.001 and $\mathcal{F}$=0.5), finding excellent agreement.

% In experiments, we observed an exponential decay of the radial displacement measured in the deformed configuration (see Fig.\,\ref{FigStretch_exp}b). We recover the exponential decay of the radial displacement with the arclength computed in the deformed configuration (plotted in dashed line). The slope of this plot defines the decay exponent which is plotted against the dimensionless axial force in Fig.\,\ref{FigStretch_theory}c. We can also measure the radial displacement  in terms of Lagrangian coordinates (i.e. using the arclegnth in the undeformed coordinates). Using this coordinate system we find that the displacement decays exponentially with an exponent scaling as $\sqrt{\F}$ (dashed line in Fig.\,\ref{FigStretch_exp}b). The experimental decay exponents are shown for three different indenter radii $R_0/a=1.46$ (blue squares), 1.95 (red circles) and 2.44 (black diamonds) and are found to agree favorably with the deformed predictions without fitting parameters. We observe a deviation at small forces that we attribute to the finite bending effect of the tube in the experiments and at very large forces that we attribute to the hyperelastic material properties of the latex at large strains.

In experiments, we observe an exponential decay of the radial
displacement with the deformed arclength $S - S_c$, which is recovered by numerical solutions of our model (symbols in Fig.\,\ref{FigStretch_exp}b). The slope of an experimental or theoretical curve on this plot defines a decay constant $\lambda$ for a given $\mathcal{F}$, corresponding to the exponential relaxation $u_r - u_{r}^{\infty} \propto \exp\left\{ -\lambda \,(S - S_c) \right\}$. The extracted $\lambda$ values, representing inverse decay lengths, are plotted against the dimensionless axial force in Fig.\,\ref{FigStretch_theory}c. Experimental decay exponents are shown for three different intruder radii $R_0/a=1.46$ (blue squares), 1.95 (red circles) and 2.44 (black diamonds) and are found to agree favorably with theoretical predictions (solid line) without fitting parameters. We find that the theoretical predictions for the decay exponents are insensitive to the intruder radius: in Fig. \ref{FigStretch_theory}, theoretical curves for different $R_0/a$ values are separated by distances smaller than the thickness of the solid black curve. Interestingly, we find that the theoretical decay exponents are equally insensitive to the presence of friction in the contact region \footnote{Frictional contact is modeled by assuming $u_z = 0$ over the contact region, which modifies \eqref{ContactGE}.}.

Axial stretching is small as $\F \rightarrow 0$, so
$S - S_c \approx s - s_c$ in this limit. Consequently, (\ref{FreeSol}a)
shows that the decay exponent becomes asymptotically
$\lambda \approx \F^{-1/2}$, which is confirmed by the dashed line in
Fig.\,\ref{FigStretch_theory}c. 
Thus, for small stretch, the tube relaxes to its cylindrical shape over a dimensional length scale $\sqrt{f a^2/(Eb)}$ or $\sqrt{F a/(2 \pi E b)}$; recall that $F$ is the applied stretching force. We observe a deviation of the theory from the experiments at very small forces that we attribute to the finite bending stiffness of the tube in the experiments, and at very large forces that we attribute to the hyperelastic material properties of the latex at large [i.e. $O(1)$] strains.

% \jens{I don't quite understand the argument here. The undeformed result
% follows trivially from (3a). The deformed result I presume is determined
% from solving the equations numerically, and then fitting an exponential?
% Is this done for zero bending rigidity? Are the two results expected to be
% the same ? Probably not, but the discussion of the experimental results
% suggests that. How were the experimental results obtained? The beginning
% of the paragraph suggests the use of deformed coordinates, the label of
% Fig.2 (b) is $s-s_c$, which suggests undeformed coordinates, because that
% is the notation that has just been used. 
% }

As discussed previously, stretching also lowers the contact area between the tube and the intruder (Fig. \ref{FigStretch_theory}a). The axial location of the contact point $z_i-z_c$ (obtained from theory)
is plotted as a function of $\F$ in Fig.\,\ref{FigvsF}. When
$\delta \equiv R_0 - 1 \ll 1$ and $\mathcal{F} \ll 1$, we find
$u_z = O(\delta^{3/2}) \ll u_r = O(\delta)$, so that $z(s) \approx s$.
Then, using a Taylor expansion of $s_c$ about $z_{i}$, we find
from \eqref{ContactCond} that $z_{i} - z_c \propto \F^{1/2}$. As
seen in Fig.\,\ref{FigvsF}, this scaling works well except for the
largest values of $\mathcal{F}$, where $z_i-z_c$ starts to depend on
$R_0$. 

We now look at the effect of bending rigidity on the regularization
of the curvature singularity by stretching. Bending rigidity of the
sheet may be included by adding a bending traction
$\sigma_b \approx \frac{E b^3}{12 (1- \nu^2)} u_r''''$ to the right side
of the radial stress balance (\ref{Shell}a) \citep{audoly_book}.
Consequently, (\ref{FreeGE}a) is modified to
$\F u_r'' - u_r - \nu \F - \beta^4 u_r'''' = 0$, where
$\beta = \left(\frac{(b/a)^{2}}{12(1- \nu^2)}\right)^{1/4}\ll 1$.  Considering the limit $\mathcal{F} \rightarrow 0$,
so that the curvature singularity is regularized by bending alone, the solution outside contact is $u_r(s>s_c) = \overline{C} \rme^{-\sqrt{\rmi}(s-s_c)/\beta} + $ c.c. The governing equations over the contact region remain unchanged from \eqref{ContactGE}. The two regions are connected by continuity of displacement ($u_r$), stress ($u_r'$) and bending moment ($u_r''$) at the contact point. For a relatively small intruder ($R_0 - 1 \ll 1$), we approximate $z \approx s$ and use these contact conditions to find $z_c \approx z_i - \sqrt{2} \beta$. Thus, the curvature-regularizing effect of stretching dominates that of bending when $\sqrt{\F} \gtrsim \sqrt{2} \beta$, or $f \gtrsim \frac{E b^2/a}{\sqrt{3 (1-\nu^2)}}$, corresponding to a linear strain of $O(b/a)$. Thus, even a small amount of axial strain can lead to the curvature of the elastic shell being dictated by stretching and not bending. 

% In our experiments, finite bending effects dominates the decay at small fo

In conclusion, we have investigated how an object is
trapped elastically inside a cylindrical tube of radius $a$ and
thickness $b$. From the competition between hoop stress and longitudinal
stretching arises a novel length scale $\sqrt{F a/(E b)}$, which
{\it decreases} with increasing sheet thickness, opposite the
usual elastic cut-off scale \cite{Witten07}, which smooths the
sheet with increasing thickness.

\begin{acknowledgments}
We are grateful to Dominic Vella for his insightful comments.
J. E. acknowledges support by the Leverhulme Trust through
International Academic Fellowship IAF-2017-010. He benefitted from
inspiring conversations with Howard Stone and his group, during an
unforgettable sabbatical year at Princeton.  B. R. and J. M. contributed equally to this work.
\end{acknowledgments}

% \bibliography{elasticity}

%

\end{document}